\documentclass[a4paper,nofootinbib]{revtex4}

\usepackage{amssymb,amsmath,graphicx}

\usepackage{color}
\usepackage[english]{babel}
\usepackage[latin1]{inputenc}
\usepackage[mathscr]{eucal}  

\usepackage{times}

\renewcommand{\hbar}{\hslash}

\newcommand{\dd}{\ensuremath{\text{d}\!\!\;}}
\newcommand{\e}{\ensuremath{\text{e}}}       
\newcommand{\HH}{\ensuremath{\mathscr H}}

\newcommand{\im}{\ensuremath{\mathrm{i}}}    
\newcommand{\GG}{{\ensuremath{\cal G}}}      
\newcommand{\GGr}{{\ensuremath{\mathrm G}}}  
\newcommand{\A}{{\ensuremath{\cal A}}}       

\newcommand{\MRS}{\ensuremath{M_{\mathrm{RS}}}}
\newcommand{\ML}{\ensuremath{M_{\mathrm{L}}}}
\newcommand{\etaS}{\ensuremath{\eta_\omega}}
\newcommand{\myvec}[1]{\vec{#1}}
\newcommand{\diag}{\ensuremath{ {\rm diag}}}
\newcommand{\Id}{{\ensuremath{\mathbf 1}}}

\newcommand{\DL}{\ensuremath{D_{\mathrm l}}}
\newcommand{\DU}{\ensuremath{D_{\mathrm u}}}
\newcommand{\NDOS}{\ensuremath{\mathscr{N}}}

\newcommand{\ket}[1]{\ensuremath{\vert{#1}\rangle}}
\newcommand{\bra}[1]{\ensuremath{\langle{#1}\vert}}

\newcommand{\average}[1]{\ensuremath{\langle{#1}\rangle}}

\newcommand{\omegaBC}{\ensuremath{\omega_{\mathrm{BC}}}}
\newcommand{\DeltaF}{\ensuremath{\Delta_{\mathrm F}}}
\newcommand{\fermi}{\ensuremath{\mathrm f}}

\newcommand{\kF}{\ensuremath{\mathrm k_{\mathrm F}}}
\newcommand{\vF}{\ensuremath{\mathrm v_{\mathrm F}}}
\newcommand{\eF}{\ensuremath{\varepsilon_{\mathrm F}}}

\definecolor{Cites}{rgb}{0.3,0,0.4}
\definecolor{White}{rgb}{1,1,1}
\definecolor{green}{rgb}{0,1,0}
\definecolor{LightRed}{rgb}{1.0,0.3,0.3}
\definecolor{darkred}{rgb}{0.6,0,0}
\definecolor{darkblue}{rgb}{0,0,0.4}
\definecolor{grey}{rgb}{0.4,0.4,0.4}

\newcommand{\Eqref}[1]{ Eq.~(\ref{#1})}
\newcommand{\Eqsref}[2]{ Eqs.~(\ref{#1}, \ref{#2})}
\newcommand{\Figref}[1]{ Figure~\ref{#1}}
\newcommand{\Tblref}[1]{ Table~(\ref{#1})}
\newcommand{\Caption}[1]{\caption{\small #1}}

\begin{document}
\title{Calculating Green Functions from Finite Systems}

\author{Peter Schmitteckert}

\affiliation{Institute of Nanotechnology, Karlsruhe Institute of Technology, 
76131 Karlsruhe Karlsruhe, Germany}

%\ead{Peter.Schmitteckert@int.fzk.de}

\begin{abstract}
In calculating Green functions for interacting quantum systems numerically
one often has to resort to finite systems which introduces a finite size
level spacing. In order to describe the limit of system size going to
infinity correctly, one has to introduce an artificial broadening
larger than the finite size level discretization.
In this work we compare various discretization schemes for impurity
problems, i.e.\ a small system coupled to leads. Starting from a naive linear
discretization we will then discuss the logarithmic discretization of the
Wilson NRG, compare it to damped boundary conditions and arbitrary discretization
in energy space. We then discuss the importance of choosing the right
single particle basis when calculating bulk spectral functions.
Finally we show the influence of damped boundary conditions on the time evolution
of wave packets leading to a NRG-tsunami.
\end{abstract}
\maketitle

\section{Introduction}
Correlation, or Green, functions are a fundamental concept of
condensed matter theory, for an introduction see e.g.\ \cite{Mahan:MPP2000,NegeleOrland:QMPS,Bruus_Flensberg:OUP2004}.
However, for interacting systems exact solutions are
rare and one often has to resort to perturbative or numerical approaches.
In calculating Green functions for infinite  systems, let it be a bulk Green function,
or an impurity problem, numerically one hast to resort to a discretized, finite system.
In this work we discuss the influence of various discretization schemes 
on the spectral function, i.e.\ the imaginary part of the retarded Green function.

To this end we restrict ourselves to non-interacting Fermi systems where numerics can be performed
without any approximation and the resulting errors can be traced back to the discretization scheme.

We start with the definition of Green functions in time domain, and derive resolvent equations in frequency domain,
which demonstrates that for the calculation of spectral functions one does not need the spectrum explicitly.
We then discuss the case of an impurity problem, namely the resonant level model, where a single level is
coupled to a non interacting lead / bath. It turns out that it is non-trivial to treat this very simple model 
accurately on a finite lattice. We then procced with the problem of an energy resolved spectral function of
a non-interacting tight binding chain. It turns out, that one can reconstruct the $\delta$-function of the spectral
function in the continuum. However, care has to be taken in order to avoid discretization errors.
Finally we show that damped, or Numerical Renormailzation Group- (NRG) like, 
boundary conditions lead to a phenomenon we call the NRG tsunami.
%%
%%%%%%%%%%%%%%%%%%%%%%%%%%%%%%%%%%%%%%%%%%%%%%%%%%%%%%%%%%%%%%%
\subsection{Green Functions in Time Domain}
%%%%%%%%%%%%%%%%%%%%%%%%%%%%%%%%%%%%%%%%%%%%%%%%%%%%%%%%%%%%%%%
%%
%%
The lesser (greater) Green functions $\GGr^<$ ($\GGr^>$) and
the retarded (advanced) Green functions $\GGr^r$ ($\GGr^a$)  are defined \cite{Mahan:MPP2000,NegeleOrland:QMPS,Bruus_Flensberg:OUP2004} by
\begin{align}
	\GGr^>_{\hat{A},\hat{B}} (t,t')	&=\;  -\im\, \average{ \hat{A}(t)  \, \hat{B}(t')} \\
	\GGr^<_{\hat{A},\hat{B}} (t,t')	&=\;  -\im  \zeta \,\average{ \hat{B}(t') \, \hat{A}(t)} \\
	\GGr^r_{\hat{A},\hat{B}} (t,t')	&=\;  - \im\, \Theta(t-t') \average{ \left[ \hat{A}(t) ,\, \hat{B}(t') \right]_{-\zeta}}  
                                 \;=\; \Theta(t-t') (\GGr^>_{\hat{A},\hat{B}} (t,t') - \GGr^<_{\hat{A},\hat{B}} (t,t')  )\label{def:CF:GreenRetarded}\\
	\GGr^a_{\hat{A},\hat{B}} (t,t')  &=\; \hphantom{-} \im\, \Theta(t'-t) \average{ \left[ \hat{A}(t),\, \hat{B}(t') \right]_{-\zeta}}
                                         \;=\; \Theta(t'-t) (\GGr^<_{\hat{A},\hat{B}} (t,t') - \GGr^>_{\hat{A},\hat{B}} (t,t')  )
\end{align}
where  $\hat{A}$ and $\hat{B}$ denote two arbitrary operators.
For fermionic operators, $\zeta=-1$, $\left[A, B\right]_+ = AB + BA$ denotes the anticommutator of operators A and B,
and for bosonic operators,  $\zeta=1$, the anticommutator is replaced by a commutator $\left[A, B\right]_- = AB - BA$.
Throughout this work $\average{\cdots}$ denotes the zero temperature ground state average.
\subsection{Resolvent Representation in Frequency Domain}
\label{ssec:CF:SPP:RRFD}
In order to simplify notation when switching to frequency domain
we assume translational invariance in time and introduce 
the following Green functions of two arbitrary operators $\hat A, \hat B$:
\begin{align}
	\GGr^+_{\hat{A},\hat{B}}(t)	&=\;           -  \im\, \Theta( t ) \, \average{ \hat{A}(t)  \, \hat{B}(0) } \\
	\GGr^-_{\hat{A},\hat{B}}(t)	&=\; \hphantom{-} \im\, \Theta( t ) \, \average{ \hat{A}(0)  \, \hat{B}(t)} \,,
\end{align}
leading to
\begin{align}
	\GGr^>_{\hat{A},\hat{B}} (t,0)	&=\; \GGr^+_{\hat{A},\hat{B}}(t) \,-\,  \GGr^-_{\hat{A},\hat{B}}(-t)\\
	\GGr^<_{\hat{A},\hat{B}} (t,0)	&=\; \zeta\GGr^+_{\hat{B},\hat{A}}(-t) \,-\, \zeta \GGr^-_{\hat{B},\hat{A}}(t)\\
	\GGr^r_{\hat{A},\hat{B}} (t,0)	&=\; \GGr^+_{\hat{A},\hat{B}}(t) \,+\,  \zeta \GGr^-_{\hat{B},\hat{A}}(t) \\
	\GGr^a_{\hat{A},\hat{B}} (t,0)  	&=\; \zeta\GGr^+_{\hat{B},\hat{A}}(-t) \,+\,  \GGr^-_{\hat{A},\hat{B}}(-t)\,. 
\end{align}
The Fourier transformed Green function $\GG^+_{\hat{A},\hat{B}}(\omega)$ is defined by
\begin{align}
	\GG^+_{\hat{A},\hat{B}}(\omega)	&=\; \int_{-\infty}^{+\infty} \dd t\,  \e^{\im \omega t } \ \GGr^+_{\hat{A},\hat{B}} (t)
	\;=\; -\im\, \int_{0}^{+\infty} \dd t\,  \e^{\im \omega t }\, \average{ \hat{A}(t)  \, \hat{B}(0)}\\
	&=\;  -\im\, \int_{0}^{+\infty} \dd t\, \e^{\im \omega t }\,  \bra{\Psi_0} \e^{\im \HH t} \,\hat{A}\, \e^{-\im \HH t} \hat{B} \ket{\Psi_0}\\
	&=\;  -\im\, \int_{0}^{+\infty} \dd t\,  \bra{\Psi_0}  \hat{A} \,\e^{\im (E_0 - \HH + \omega + \im\eta) t } \, \hat{B} \ket{\Psi_0}\\
	&=\;  -\im\, \left. \bra{\Psi_0}  \hat{A} \,\frac{\e^{\im (E_0 - \HH + \omega + \im\eta)t }}{\im (E_0 - \HH + \omega + \im\eta) }\right|^\infty_{0} \, \hat{B} \ket{\Psi_0} \\
	&=\;  \bra{\Psi_0}  \hat{A} \,\frac{1}{ E_0 - \HH + \omega + \im\eta }  \, \hat{B} \ket{\Psi_0} \label{eq:CF:GeneralResolventExpressionGp}
\end{align}
and similarly
\begin{align}
	\GG^-_{\hat{A},\hat{B}}(\omega)	&=\;  \bra{\Psi_0}  \hat{A} \,\frac{1}{ E_0 - \HH - \omega - \im\eta }  \, \hat{B} \ket{\Psi_0} \,,\label{eq:CF:GeneralResolventExpressionGm}
\end{align}
where a convergence generating $\eta=0^+$ has been introduced to ensure convergence.
%
% % % Note that
% % % \begin{align}
% % % 	\GG^+_{\hat{A},\hat{B}}(-\omega)	&=\; \int_{-\infty}^{+\infty} \dd t\,  \e^{-\im \omega t } \ \GGr^+_{\hat{A},\hat{B}} (t)
% % % 		\;=\; \int_{-\infty}^{+\infty} \dd t\,  \e^{ \im \omega t } \ \GGr^+_{\hat{A},\hat{B}} (-t) \\
% % % 	\GGr^+_{\hat{A},\hat{B}}(-t)	&=\;           -  \im\, \Theta( -t ) \, \average{ \hat{A}(-t)  \, \hat{B}(0) } \\
% % %            	&=\;   -  \im\, \Theta(-t ) \, \bra{\Psi_0} \e^{-\im \HH t} \,\hat{A}\, \e^{\im \HH t} \hat{B} \ket{\Psi_0} 
% % % 		\;=\;  -  \im\, \Theta(-t ) \, \bra{\Psi_0}\hat{A}\, \e^{\im \HH t} \hat{B}   \, \e^{-\im \HH t}\ket{\Psi_0}  \\
% % % 		&=\;   -  \im\, \Theta(-t ) \, \average{ \hat{A}(0)  \, \hat{B}(t) } \\
% % % 	\GGr^-_{\hat{A},\hat{B}}(-t)	&=\;             \im\, \Theta( -t ) \, \average{ \hat{A}(0)  \, \hat{B}(-t) } 
% % % 		\;=\;     \im\, \Theta(-t ) \, \average{ \hat{A}(t)  \, \hat{B}(0) } \,.
% % % \end{align}
%
Finally, we obtain
\begin{align}
	\GG^>_{\hat{A},\hat{B}} (\omega)	&= \GG^+_{\hat{A},\hat{B}}(\omega) \,-\, \hphantom{\zeta}\GG^-_{\hat{A},\hat{B}}(-\omega) \\
	\GG^<_{\hat{A},\hat{B}} (\omega)	&= \zeta\GG^+_{\hat{B},\hat{A}}(-\omega) \,-\, \zeta \GG^-_{\hat{B},\hat{A}}(\omega) \\
	\GG^r_{\hat{A},\hat{B}} (\omega)	&= \GG^+_{\hat{A},\hat{B}}(\omega) \,+\, \zeta\GG^-_{\hat{B},\hat{A}}( \omega) \\
	\GG^a_{\hat{A},\hat{B}} (\omega)	&=  \zeta\GGr^+_{\hat{B},\hat{A}}(-\omega) \,+\,  \GGr^-_{\hat{A},\hat{B}}(-\omega)
\end{align}

%%%%%%%%%%%%%%%%%%%%%%%%%%%%%%%%%%%%%%%%%%%%%%%%%%%%%%%%%%%%%%%
\subsection{Correction Vector Approach}
%%%%%%%%%%%%%%%%%%%%%%%%%%%%%%%%%%%%%%%%%%%%%%%%%%%%%%%%%%%%%%%
%%
It is interesting to note that the resolvent representation \Eqsref{eq:CF:GeneralResolventExpressionGp}{eq:CF:GeneralResolventExpressionGm}
allows for calculating the Green function without a complete knowledge of the spectrum via the correction vector approach \cite{Ramasesha:JCC1990}.
Starting from a general resolvent expression
\begin{equation}
	G(E) = \bra{ \Psi} \hat{A} \underbrace{\frac{ 1}{ H - E + \im \eta } \hat{B} \ket{ \Psi}}_{\displaystyle \ket{\xi}}
			\;=\;   \bra{ \Psi} \hat{A}\ket{\xi}
\end{equation}
we obtain the correction vector $\ket{\xi}$ from the linear system
\begin{equation}
\left(H - E +  \im \eta\right) \ket{\xi} = \hat{B} \ket{ \Psi} \,,
\end{equation}
which can be solved by standard solvers. However, note that in this approach one needs a separate run for each
desired energy $E$.

%%%%%%%%%%%%%%%%%%%%%%%%%%%%%%%%%%%%%%%%%%%%%%%%%%%%%%%%%%%%%%%
\subsection{Single Particle Propagator}
%%%%%%%%%%%%%%%%%%%%%%%%%%%%%%%%%%%%%%%%%%%%%%%%%%%%%%%%%%%%%%%
%%
The single particle Green functions for fermionic systems are defined by
\begin{align}
	\GGr^>(x,t; y,t')	&=\;  -            \im\, \average{ c^{}_x(t)  \, c^+_y(t')} \\
	\GGr^<(x,t; y,t')	&=\;  \hphantom{-} \im\, \average{ c^{+}_y(t') \, c^{}_x(t)} \\
	\GGr^r(x,t; y,t')	&=\;  -            \im\, \Theta(t-t') \average{ \left[ c^{}_x(t)  ,\, c^+_y(t') \right]_{+}} \label{def:SPP:GreenRetarded}\\
	\GGr^a(x,t; y,t')	&=\;  \hphantom{-} \im\, \Theta(t'-t) \average{ \left[ c^{}_x(t),\, c^{+}_y(t') \right]_{+}}  
\end{align}
where  we use $x$ and $y$ to denote the position in the lattice, $c^{}_x(t)$ and $c^+_x(t)$ are the 
fermionic annihilation and creation operators at site $x$ and time $t$. For spinful calculations
$x$ denotes a super index of the spatial coordinate and the spin orbital. 
From Eqs.~(\ref{eq:CF:GeneralResolventExpressionGp}, \ref{eq:CF:GeneralResolventExpressionGm}) we obtain for the retarded and advanced Green functions
\begin{align}
	\GG^r(x,y,\omega)	&=\; \GG^+_{ \hat{c}^{}_x,\hat{c}^+_y}(\omega) \;-\;  \GG^-_{ \hat{c}^{+}_y,\hat{c}^{}_x}(\omega) \label{eq:CF:RetartedRRFD}\\
	\GG^a(x, y, \omega)	&=\; \GG^r(x,y,\omega)^* \,. \label{eq:CF:AdvancedRRFD}
\end{align}
%%
% % From Eqs.~(\ref{eq:CF:GeneralResolventExpressionGp}, \ref{eq:CF:GeneralResolventExpressionGm}) we obtain the single particle propagators
% % \begin{align}
% % 	\GG^>(x,y,\omega)	&=\; \GG^+_{ \hat{c}^{}_x,\hat{c}^+_y}(\omega) \,-\, \hphantom{\zeta}\GG^-_{\hat{c}^{}_x,\hat{c}^+_y}(-\omega)\\
% % 	\GG^<(x,y,\omega)	&=\;\zeta\GG^+_{\hat{c}^+_y,\hat{c}^{}_x}(-\omega) \,-\, \zeta \GG^-_{\hat{c}^+_y,\hat{c}^{}_x}(\omega) \\
% % 	\GG^r(x,y,\omega)	&=\; \GG^+_{ \hat{c}^{}_x,\hat{c}^+_y}(\omega) \;+\; \zeta\, \GG^-_{ \hat{c}^{+}_y,\hat{c}^{}_x}(\omega) \label{eq:CF:RetartedRRFD}	\\
% % 			&=\;  \bra{\Psi_0}  \hat{c}^{}_x \,\frac{1}{ E_0 - \HH + \omega + \im\eta }  \, \hat{c}^{+}_y \ket{\Psi_0}
% %    			\;+\; \zeta\, \bra{\Psi_0}  \hat{c}^{+}_y \,\frac{1}{ E_0 - \HH - \omega - \im\eta }  \, \hat{c}^{}_x \ket{\Psi_0} 	\\
% % 	\GG^a(x, y, \omega)	&=\; \GG^r(x,y,\omega) ^* \,. \label{eq:CF:AdvancedRRFD}
% % \end{align}
%
%%%%%%%%%%%%%%%%%%%%%%%%%%%%%%%%%%%%%%%%%%%%%%%%%%%%%%%%%%%%%%%
\subsection{Free Fermions}
%%%
Up to now the description for Green functions was completely general.
In the following we restrict ourselves to the description of non-interacting Fermi systems
since our goal is to describe the problem induced by evaluating Green functions on finite systems.
In result we are able to perform approximation free numerics. Nevertheless, our findings are
applicable for interacting system and can be exploited by other methods.

For non-interacting fermions we start with a general Hamiltonian
\begin{equation}
	\HH \;=\; \myvec{\hat{c}}^+  \cdot H \cdot \myvec{\hat{c}} \;=\; \sum_{x,y}  \hat{c}^+_x \, H_{x,y} \, \hat{c}^{}_y \,.
\end{equation}
We can now switch to a diagonal basis by diagonalizing the matrix $H$:
\begin{align}
	\diag( \myvec{\varepsilon}) &=  U \cdot H \cdot U^+ \\
	\Id                         &=  U \cdot U^+ \\
	\myvec{\tilde{c}}           &=  U \cdot  \myvec{\hat c} \,,
\end{align}
where $\varepsilon_\ell$ are the single particle levels.
If the ground state $\ket{\Psi_0}$ is non degenerate it is given by
\begin{align}
	\ket{\Psi_0} &=  \prod_{\varepsilon_\ell < \eF} \tilde{c}^+_\ell \ket{-} \,,
\end{align}
where \eF\ is the Fermi energy and $\ket{-}$ is the vacuum state.
However,
on finite systems at zero temperature this definition of the Fermi energy is ambiguous
since \eF\ can sit anywhere between the highest occupied and the lowest
unoccupied level. 
We set \eF\ in the middle of those two levels to ensure numerical stability.
For degenerate ground states one has to take care of the different possibilities of filling the highest level.
When evaluating expectation values one then has to average degenerate levels at the Fermi energy
by taking the zero temperature limit of the finite temperature result, e.g.\ the number of particles $N$ 
is then given by
\begin{align}
	N &= \lim_{\beta \rightarrow  \infty} \sum_\ell \fermi\left( \beta ( \varepsilon_\ell -\eF) \right) \, U^*_{\ell,x} U^{}_{\ell,x} \,,
\end{align}
where $\fermi()$ is the fermi function and one should work at a small, but non-vanishing temperature.
\par
Evaluating the retarded single particle Green functions 
$\GGr^r( x,t;y,t' )$
of \Eqref{eq:CF:RetartedRRFD} in frequency domain using the formulae of section \ref{ssec:CF:SPP:RRFD} we obtain
\begin{align}
	\GG^+_{\hat{c}_x,\hat{c}^+_y}(\omega)	&=  \bra{\Psi_0}  \hat{c}^{}_x \,\frac{1}{ E_0 - \HH + \omega + \im\eta }  \, \hat{c}^+_y \ket{\Psi_0} 
	=  \sum_{\ell} \left( 1 - \bra{\Psi_0}\tilde{n}_\ell\ket{\Psi_0}\right)  \,\frac{ U^*_{\ell,x}\, U^{}_{\ell,y}}{ \omega  -\varepsilon_\ell + \im\eta } 
% % % 	\\
% % % 	&=  \sum_{\ell,m} \bra{\Psi_0}  \tilde{c}^{}_\ell \,\frac{ U^+_{x,\ell}\, U^{}_{m,y}}{ E_0 - (E_0+\varepsilon_m) + \omega + \im\eta }  \, \tilde{c}^+_m \ket{\Psi_0} \\
% % % 	&=  \sum_{\ell} \left( 1 - \bra{\Psi_0}\tilde{n}_\ell\ket{\Psi_0}\right)  \,\frac{ U^*_{\ell,x}\, U^{}_{\ell,y}}{ \omega  -\varepsilon_\ell + \im\eta } 
\end{align}
and 
\begin{align}
	\GG^-_{\hat{c}^+_y,\hat{c}^{}_x}(\omega)	&=\;  -\sum_{\ell}  \bra{\Psi_0}\tilde{n}_\ell\ket{\Psi_0} \,\frac{ U^{*}_{\ell,x}\, U^{}_{\ell,y}}{ \omega  -\varepsilon_\ell + \im\eta } \,.
\end{align}
Finally we obtain from \Eqsref{eq:CF:RetartedRRFD}{eq:CF:AdvancedRRFD}
\begin{align}
\GG^r(x,y,\omega)	&=\; \GG^+_{ c^{}_x,c^+_y}(\omega) \,-\, \GG^-_{ c^{+}_y,c^{}_x}(\omega)
	\;=\; \sum_{\ell} \frac{ U^{*}_{\ell,x}\, U^{}_{\ell,y}}{ \omega - \varepsilon_\ell  + \im\eta } \label{eq:FFL:RetartedRRFD} 
% % \GG^a(x,y,\omega)	&=\; \GG^r(x,y,\omega)^*
% % 	\;=\; \sum_{\ell} \frac{ U^{}_{\ell,x}\, U^{*}_{\ell,y}}{ \omega - \varepsilon_\ell  - \im\eta } \label{eq:FFL:AdvancedRRFD} \\
% % \GG^>(x,y,\omega)	&=\; \GG^+_{ c^{}_x,c^+_y}(\omega) \,-\, \GG^-_{ c^{}_x,c^{+}_y}(-\omega)\nonumber\,.
\end{align}
%%
%%
%%%%%%%%%%%%%%%%%%%%%%%%%
\section{Resonant Level Model}
%%%%%%%%%%%%%%%%%%%%%%%%%
Following the rather general introduction on calculating Green functions we will now concentrate
on the spectral function
\begin{align}
	\A &= \frac{-1}{\pi} \Im\, \GG^r(x,x,\omega)
\end{align}
 of a single resonant level $\epsilon_d \hat{n}_d$ coupled via a hybridization
$t'$ to a onedimensional lead , where we set the hopping element to $t=1$.
Ignoring the finite width and the cosine dispersion of the lead band results in the wide band limit solution
of an area normalized Lorentzian
%%%	\A &= \frac{1}{\pi} \frac{w}{ \omega^2 + w^2}  \qquad w = \frac{2 \tL^2 \tR^2}{\tL^2 + \tR^2} \label{eq:CF:AofRLM}
\begin{align}
	\A &= \frac{1}{\pi} \frac{w}{ \omega^2 + w^2}  \qquad w = t'^2 \label{eq:CF:AofRLM}
\end{align}
Throughout this section we apply \Eqref{eq:FFL:RetartedRRFD}
to evaluate the resolvent equation \eqref{eq:CF:RetartedRRFD}.
A problem that arises is that the convergence factor $\eta$ of \eqref{eq:CF:RetartedRRFD}, which  is $0^+$ for continuum leads
has to be larger than the finite size level splitting of the leads while it has to be much smaller than any
physical scale of interest as it also acts as a broadening. In addition, in finite systems 
the existence of a boundary influences the ground state result typically on a scale $ \omegaBC \sim \DeltaF$,
where \DeltaF\ is the level spacing at the Fermi surface.
Note, that in the case of a single level coupled to a lead with finite width there may exist
bound states outside the conduction band. While these states are not accessible by single particle
of the conduction band due to energy conservation, they can be access by few- or many- particle 
processes.\cite{Longo_Schmitteckert_Busch:PRL2010}.

The goal of this section is to provide an overview of various discretization schemes of the leads 
and their impact on the spectral function. 
We start with the natural choice of a finite nearest neighbour tight binding chain.
It turns out that their resolution is quiet limited.
Next we discuss
discretization schemes used in the Numerical Renormalization Group (NRG) approach\cite{Krishnamurthy_Wilkins_Wilson:PRB1980}
and a variation
of it called Smooth\cite{Vekic_White:PRB1993} or Damped\cite{BohrSchmitteckert:PRB2007} Boundary conditions, 
which give a good result for low frequencies, but the high
frequency results are spoiled. Finally we provide a discretization scheme which is able to
provide high resolution on all energy scales, where we generalize the variable discretization approach of
Nishimoto and Jeckelmann.\cite{Nishimoto_Jeckelmann:JPCM:2004}
%%
%%%%%%%%%%%%%%%%%%%%%%%%%
\subsection{Linear Leads}
\begin{figure}[ht]
\begin{center}
	\includegraphics[width=0.9\textwidth]{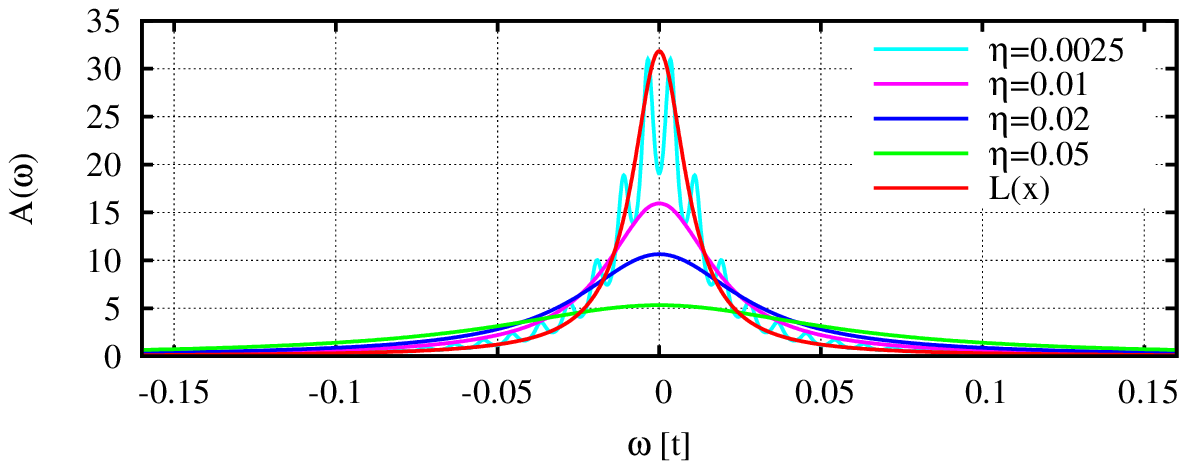}
\end{center}
\Caption{Spectral function of a resonant level, $t'= 0.1$, $M=700$, $N=350$, $\epsilon_d=0$, HWBC,
and $\eta=0.05$, 0.02,  0.01, 0.0025. $L(\omega)$  shows the Lorentzian  of half halfwidth $w=0.01$ corresponding to the spectral function
in the wide band limit.} \label{fig:CF:NEFS:SF_Linear}
\end{figure}

%%%%%%%%%%%%%%%%%%%%%%%%%
%%
In \Figref{fig:CF:NEFS:SF_Linear} we show the numerical evaluation of the spectral function of  a resonant level  coupled 
to a lead via a hybridization of $t' = 0.1$ evaluated for a total number of 700 sites and 350 fermions and hard wall
boundary conditions (HWBC).
%%
%%
%%--------------------------------------------------------------------
\begin{figure}[ht]
\begin{center}
	\includegraphics[width=0.9\textwidth]{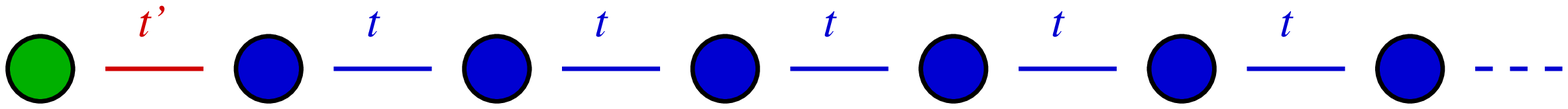}
\end{center}
\Caption{Sketch of a single impurity coupled to a lead via $t'$ and a lead hopping of $t$.} \label{fig:CF:NEFS:Fig_SF_Linear}
\end{figure}
%%--------------------------------------------------------------------
%%
%%
The corresponding Lorentzian \eqref{eq:CF:AofRLM} has a half halfwidth of $w = t'^2$=0.01 which is already
larger than the finite size level splitting of $\Delta_F\approx 0.00352$.
%%
%%--------------------------------------------------------------------
%%--------------------------------------------------------------------
%%
%% 
Nevertheless the figure demonstrates that the discretization  of the leads is too coarse to reproduce
the Lorentzian  although the level spacing is already significantly smaller than the resonance width. 
For an $\eta$  larger than the finite size level spacing the resonance is artificially broadened while
for smaller $\eta$ the discrete nature of leads appears in the correlation function as can be seen by the spikes of
the $\eta=0.0025$ result.  This observation is explained by looking more closely at the resolvent equation \eqref{eq:CF:RetartedRRFD}.
In the thermodynamic limit of continuum leads $\eta$ is $0^+$ and the imaginary part of the resolvent $\GG^+_{ c^{}_x,c^+_y}(\omega)$
\begin{align}
	\Im\, \frac{1/\pi}{ E_0 - \HH + \omega + 0^+} &= \pi \delta( \omega - (\HH-E_0) )  
\end{align}
gives a contribution only at $\omega = E_n - E_0$, where $E_n$ are the eigenenergies of the Hamiltonian \HH. 
By switching to a finite $\eta$ the $\delta$ function gets replaced by 
a Lorentzian of half halfwidth $\eta$. This corresponds to a convolution of the original spectral function with a Lorentzian
of the same width and our result is replaced by
\begin{align}
	L_\eta(\omega) &= \int_{-\infty}^{\infty} \dd \varepsilon \,  \frac{ w /\pi}{ \left(\omega-\varepsilon\right)^2 + w^2 }  \, \frac{ \eta /\pi }{ \varepsilon^2 + \eta^2 }  
	\;=\;\frac{1}{\pi} \frac{ w + \eta }{ \omega^2  + ( w + \eta)^2 } \,.\label{eq:CF:ConvolutedLorentzian}
\end{align}
%%
%%------------------------------------------------------
\begin{figure}[ht]
\begin{center}
	\includegraphics[width=0.9\textwidth]{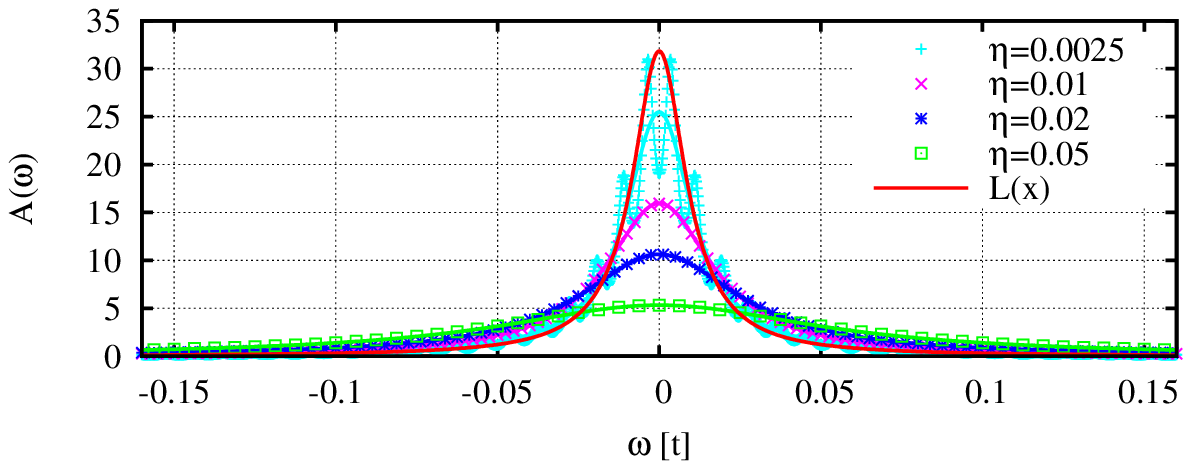}
\end{center}
\Caption{Lorentzian fits of the spectral function of a resonant level, $t'=0.1$, $M=700$, $N=350$, $\epsilon_d=0$, HWBC,
and $\eta=0.05$, 0.02,  0.01, 0.0025. The level spacing at the Fermi energy is $\Delta_F\approx 0.00352$.
The full line shows the Lorentzian  of half halfwidth $w=0.01$ corresponding to the spectral function
in the wide band limit.} \label{fig:CF:NEFS:SF_LinearFit}
\end{figure}
%%----------------------------------------------------------
%%
By taking into account the broadening induced by $\eta$ we can fit \Eqref{eq:CF:ConvolutedLorentzian} to our data
as shown in \Figref{fig:CF:NEFS:SF_LinearFit} and Table~\ref{tbl:CF:NEFS:SF_Fit}. The rather good results for even large $\eta$
suggest that the results may be strongly improved by unfolding the $\eta$ broadening by a deconvolution.
\begin{table}[htb]
\begin{center}
\begin{tabular}{|l|l|l|l|l|l|l|} \hline\hline
	$\eta$                 & 0.1     & 0.05     &  0.02    & 0.01    & 0.0025  &  0.001   \\\hline
      $w_{\mathrm L}$        & 0.00942 & 0.00973  &  0.00990 & 0.00995 & 0.00999 &  0.0170  \\
      $w_{\mathrm{Log}}$     & 0.00855 & 0.00876  &  0.00887 & 0.00891 & 0.00894 &  0.00895  \\
      $w_{\mathrm{DBC}}$     & 0.00945 & 0.00974  &  0.00987 & 0.00979 & 0.00942 &  0.00929 \\
	\hline\hline
\end{tabular}
\end{center}
\Caption{Fits of a Lorentzian  \eqref{eq:CF:ConvolutedLorentzian} to the spectral function of a single impurity coupled to
a single lead. 
All results are in units of the lead hopping element $t$. $w_{\mathrm L}$ corresponds to the linear
leads used in \Figref{fig:CF:NEFS:SF_Linear} and  fits are shown in \Figref{fig:CF:NEFS:SF_LinearFit}.
$w_{\mathrm{Log}}$ corresponds to the logarithmic discretization used in \Figref{fig:CF:NEFS:Fig_SF_Log}
and $w_{\mathrm{DBC}}$ corresponds to the damped boundary conditions of \Figref{fig:CF:NEFS:Fig_SF_DBC}
The analytical result is $w=0.01t$.}  \label{tbl:CF:NEFS:SF_Fit}
\end{table}
%%
%%
%%
%%%%%%%%%%%%%%%%%%%%%%%%%%%%%%%%%%%%%%%%
\subsection{Logarithmic Discretization}
\label{sec:CF:NEFS:LD}
\begin{figure}[ht]
\begin{center}
	\includegraphics[width=0.9\textwidth]{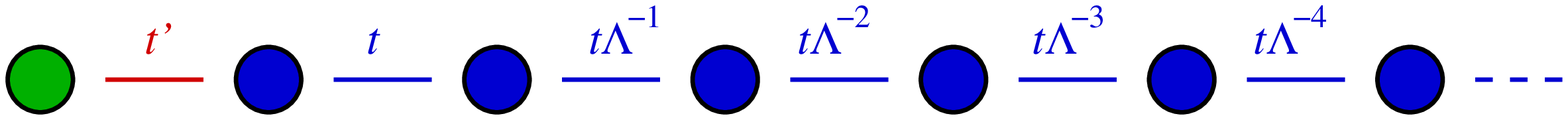}
\end{center}
\Caption{Sketch of a single impurity coupled to a lead via $t'$ and a NRG like lead hopping of $t \Lambda^{-n}$.} \label{fig:CF:NEFS:Fig_SF_Log}
\end{figure}

%%%%%%%%%%%%%%%%%%%%%%%%%%%%%%%%%%%%%%%%
%%
In order to increase the energy resolution of our lead we can replace the tight binding
chain with constant hopping by a tight binding chain similar to the one used in NRG where the hopping element
is exponentially decreased by a factor $\Lambda^{-n/2}$, with $\Lambda>1$ and $n$ the index of the NRG iteration.\cite{Krishnamurthy_Wilkins_Wilson:PRB1980}
Here we use a chain where the hopping is reduced by a factor of $\Lambda$ on each bond as displayed in \Figref{fig:CF:NEFS:Fig_SF_Log}.
\par
%%
%%--------------------------------------------------------------------
\begin{figure}[ht]
\begin{center}
	\includegraphics[width=0.9\textwidth]{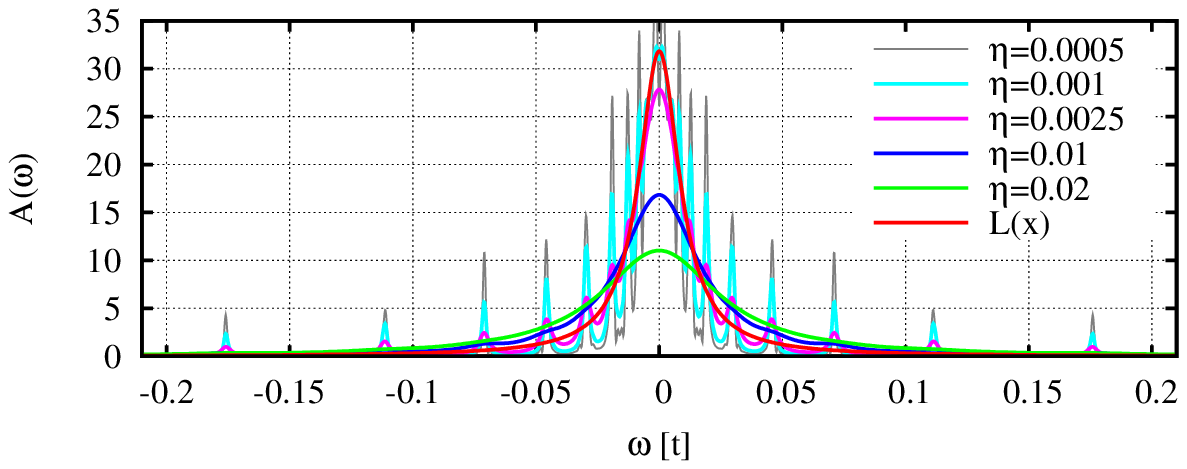}
\end{center}
\Caption{Spectral function of a resonant level, $\epsilon_d=0$, coupled to 31 NRG like lead sites via $t'= 0.1 t$
using $t=1$, $\Lambda^{-1}=0.6$. The finite size level spacing is  $\DeltaF\approx 0.000739t$.
$L(\omega)$ shows the Lorentzian  of half halfwidth $w=0.01$ corresponding to the spectral function
in the wide band limit.}\label{fig:CF:NEFS:SF_Log}
\end{figure}

%%--------------------------------------------------------------------
%%
In \Figref{fig:CF:NEFS:SF_Log} we show the results for a single level coupled to 31 NRG like lead sites
where the hopping $t=1$ is reduced by $\Lambda^{-1}=0.8$ on each bond as sketched in \Figref{fig:CF:NEFS:Fig_SF_Log}.
For these parameters we get a levelspacing at the Fermi surface of $\DeltaF\approx 0.000739t$. Correspondingly the resolution
for $\omega=0$ is now much higher and the $\eta=0.0025$ curve is well resolved at small frequencies.
However, for large frequencies the resolution drops exponentially leading to spikes even in the $\eta=0.01t$ curve.
As a consequence the Lorentzian fits do not work as well as with the (albeit much larger) linear lead, compare
\Tblref{tbl:CF:NEFS:SF_Fit}. A general feature of the logarithmic discretization consists in the overly
broad tails, compare also the the section on frequency dependent broadening.
%%
%%--------------------------------------------------------
%%--------------------------------------------------------
%%
%%
%%%%%%%%%%%%%%%%%%%%%%%%%
\subsection{Damped Boundary Conditions}
\begin{figure}[ht]
\begin{center}
\includegraphics[width=0.9\textwidth]{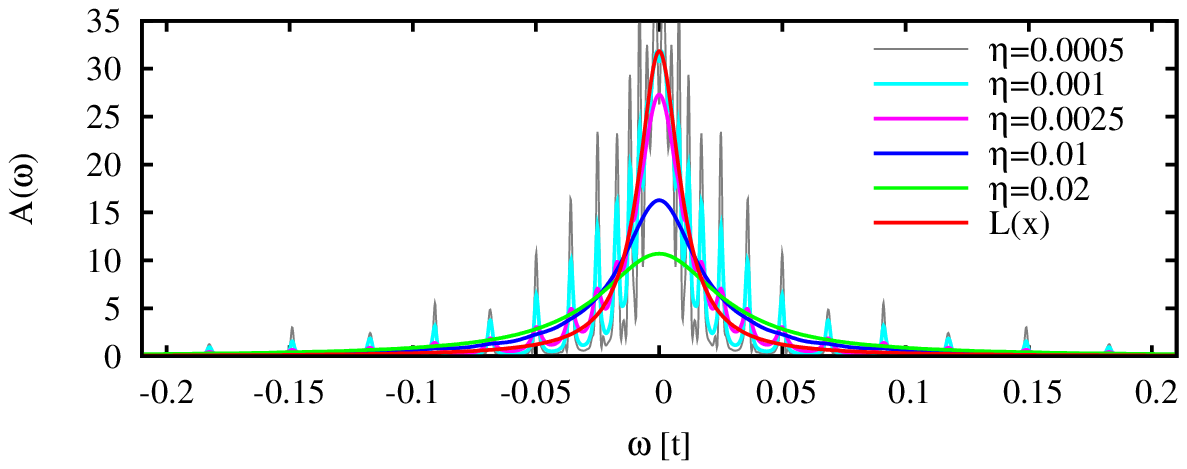}
\end{center}
\Caption{Spectral function of a resonant level, $\epsilon_d=0$, coupled to an NRG like Hamiltonian via $t'= 0.1$
as in \Figref{fig:CF:NEFS:SF_Log},
$\Lambda^{-1}=0.8$, $M_\Lambda=30$, plus 100 additional sites with a constant hopping of $t$ inserted between
the impurity and the NRG like chain leading to a level spacing at the Fermi surface of $\DeltaF\approx 0.00144$. 
The fits of Lorentzian \eqref{eq:CF:ConvolutedLorentzian} are shown in \Tblref{tbl:CF:NEFS:SF_Fit}.
The full red line shows the Lorentzian  of half halfwidth $w=0.01$ corresponding to the spectral function
in the wide band limit.} \label{fig:CF:NEFS:SF_DBC}
\end{figure}

%%%%%%%%%%%%%%%%%%%%%%%%%
%%
% In this section we combine the linear and exponentially decaying leads by first attaching the
impurity to $M_{\mathrm{RS}}$ real space sites with constant hopping element $t$ and then coupling to
an exponentially decaying lead. 
%%
%%
%%--------------------------------------------------------------------
\begin{figure}[htb]
\begin{center}
	\includegraphics[width=0.9\textwidth]{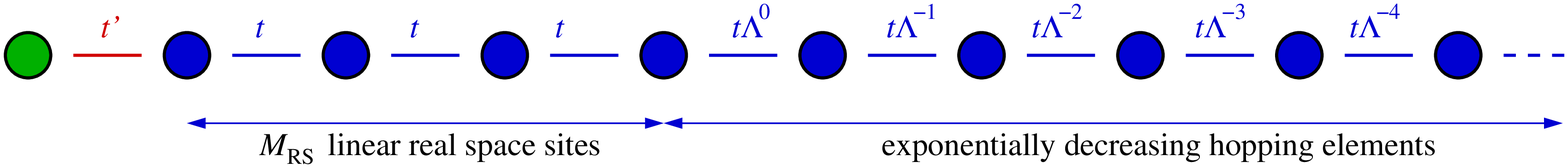}
\end{center}
\Caption{Sketch of a single impurity coupled to a lead via $t'$ and a NRG like lead hopping of $t \Lambda^{-n}$.} \label{fig:CF:NEFS:Fig_SF_DBC}
\end{figure}
%%--------------------------------------------------------------------
In \Figref{fig:CF:NEFS:SF_DBC} we show the result for a setup that corresponds to \Figref{fig:CF:NEFS:SF_Log}
where a hundred additional sites with hopping element $t$ have been inserted between the impurity and exponential decaying lead.
A similar kind of boundary condition has been originally introduced by Veki\'{c} and White\cite{Vekic_White:PRB1993} 
to mimic the thermodynamic limit in a bulk system.
The version employed here was introduced by {Bohr}, {Schmitteckert}, and {Wï¿½lfle}  \cite{BohrSchmitteckertWoelfle:EPL2006} to tackle the finite
size effects in the evaluation of the Kubo formula for linear transport. 
%%
%%----------------------------------------------
%%-------------------------------------------------
%%
At a first glance the additional lead sites result in a slight reduction of the spike of the logarithmic
discretization only. However, one obtains significantly better Lorentzian \eqref{eq:CF:ConvolutedLorentzian} fits compared
to the Logarithmic discretization alone. Nevertheless, the result is still quite disappointing.
%%
%%%%%%%%%%%%%%%%%%%%%%%%%%%%%%%%%%%%%%%%%%%
\subsection{Frequency Dependent Broadening}
\begin{figure}[ht]
\begin{center}
\includegraphics[width=0.9\textwidth]{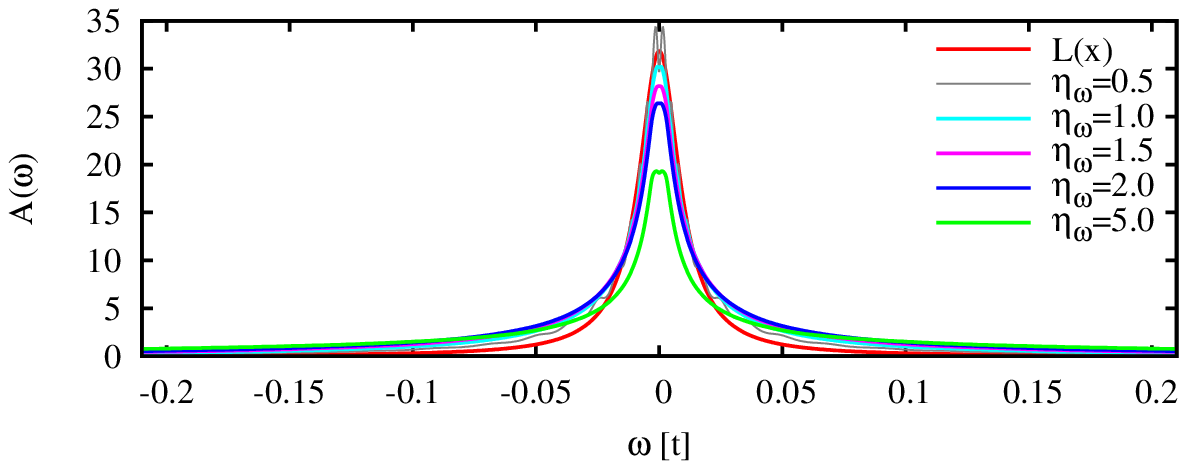}
\end{center}
\Caption{Spectral function of a resonant level using DBC as in \Figref{fig:CF:NEFS:SF_DBC}, 
$\epsilon_d=0$, $t'= 0.1$, $\Lambda^{-1}=0.8$, $M_\Lambda=30$, $\MRS=30$, and  $\DeltaF\approx 0.00144$. 
Here we used a broadening $\eta = \etaS * \Delta_\omega$, where $\Delta_\omega$ corresponds to the level spacing
at energy $\omega$, see \Eqref{CF:DeltaOmega}.
The full red line shows the Lorentzian  of half halfwidth $w=0.01$ corresponding to the spectral function
in the wide band limit.} \label{fig:CF:NEFS:SF_FDB}
\end{figure}

%%%%%%%%%%%%%%%%%%%%%%%%%%%%%%%%%%%%%%%%%%%
%%%%
%%
In order to remove the spikes in the logarithmic or DBC discretization one has to employ frequency dependent broadening.
In \Figref{fig:CF:NEFS:SF_FDB} we present the results for the same system as in \Figref{fig:CF:NEFS:SF_DBC}.
However, this time we employ a broadening $\eta$ that is proportional to the level spacing at energy $\omega$. Since
the energy spectrum is discrete we take a linear weighted average of the level spacing of the level below and above $\omega$.
Let us define the level spacing $\Delta_\omega $ at energy $\omega$ as
\begin{align}
	\Delta_\omega = ( \varepsilon_{n} -  \varepsilon_{n-1} )  
                    +  (\varepsilon_{n+1} - 2 \varepsilon_{n} + \varepsilon_{n-1} ) \frac{ \omega - 0.5\, ( \varepsilon_{n} + \varepsilon_{n-1} )}{ 0.5\, (\varepsilon_{n+1} - \varepsilon_{n-1}) }
	\label{CF:DeltaOmega}
\end{align}
where $n$ is  the level index with
\begin{align}
	0.5 \,( \varepsilon_{n-1} +  \varepsilon_{n} ) \;\le\; \omega \;<\;  0.5 \,( \varepsilon_{n} +  \varepsilon_{n+1} )  \,.
\end{align}
Note, that in the case of degenerate levels one should only count distinct energy levels to avoid a vanishing distance.
For an $\omega$ outside the energy range where the corresponding $n$ exists we use the level spacing of the corresponding
first or last level distance.
We then define a relative $\eta$ scaling 
\begin{align}
	\eta &= \eta_\omega * \Delta_\omega \,.
\end{align}
For an $\eta_\omega=1$ we obtain a broadening $\eta$ that is of the order of the level spacing at  energy $\omega$.
In \Figref{fig:CF:NEFS:SF_FDB} we compare spectral functions for the same system as in \Figref{fig:CF:NEFS:Fig_SF_DBC},
only the constant $\eta$ is replaced by a relative $\eta$ broadening.
With this approach one can eliminate the spikes by using an $\eta_\omega \ge 1.0$. 
However, the tails are still too broad and \Eqref{eq:CF:ConvolutedLorentzian} can not be used to fit the result,
as $\eta$ is now energy dependent. In order to filter the $\eta$ induced broadening one would now have to resort to
an energy dependent deconvolution.
%%----------------------------------------------
%%-------------------------------------------------
%%
%%
%%%%%%%%%%%%%%%%%%%%%%%%%
\subsection{Frequency Adapted Grids}
%%%%%%%%%%%%%%%%%%%%%%%%%
%%
In the previous sections we showed different lattice schemes to evaluate a spectral function.
None of the schemes gave actually satisfying results. Either the resolution was poor or the tails
were not represented correctly. A solution to this problem consists in adapting the lattice
for each frequency $\omega$.

We would like to note  that this is a generalization to the variable discretization approach 
of Nishimoto and Jeckelmann\cite{Nishimoto_Jeckelmann:JPCM:2004} in the sense that we can apply a constant
broadening for the complete frequency range due to our recipe for constructing discretizations.
%%
%%
%%%%%%%%%%%%%%%%%%%%%%%%%
\subsubsection{Momentum Space Leads}
%%%%%%%%%%%%%%%%%%%%%%%%%
%%
This goal can be achieved by switching to leads in momentum (or energy) space.
Let us start with the infinite chain\footnote{Strictly speaking, %
we should replace the finite lead by a semi-infinite chain. However, this could be
incorporated into $t_x$.}
\begin{align}
 \sum_{x=0}^{M} t^{}_x \hat{c}^{+}_{x} \hat{c}^{}_{x-1} \,+\, t^{*}_x \hat{c}^{+}_{x-1} \hat{c}^{}_{x} &\rightarrow\; 
	\sum_{x=-\infty}^{\infty} t^{}_x \hat{c}^{+}_{x} \hat{c}^{}_{x-1} \,+\, t^{*}_x \hat{c}^{+}_{x-1} \hat{c}^{}_{x}  \,.
\end{align}
We now switch to momentum representations
\begin{align}
 \sum_{x=-\infty}^{\infty} t^{}_x \hat{c}^{+}_{x} \hat{c}^{}_{x-1} \,+\, t^{*}_x \hat{c}^{+}_{x-1} \hat{c}^{}_{x} 
	&=\; \frac{1}{2\pi} \int_{-\pi}^{\pi} \dd k \,\epsilon^{}_k\, \tilde{c}^{+}_{k} \tilde{c}^{}_{k} \,.
\end{align}
For a nearest neighbour chain one obtains
$\epsilon(k) = -2t \cos(k)$, however one can now use any desired band $\epsilon(k)$.
Motivated by these considerations we use the general form of a lead in 'momentum space'
\begin{align}	 
	 \frac{1}{2\pi} \int_{\DL}^{\DU} \dd k \, \NDOS(k)\,\epsilon^{}_k\, \tilde{c}^{+}_{k} \tilde{c}^{}_{k} \,,
\end{align}
where $ \NDOS(k)$ is the momentum density of states, and
\DL\ ( \DU) the lower (upper) momentum cutoff. Note, here we name '$k$' momentum although it can
be any labelling. We avoid using an energy density of states since in this work as we might be interested in transport properties
and keep the flexibility to describe left and right movers by negative and positive momenta.
%%
%%%%%%%%%%%%%%%%%%%%%%%%%
\subsubsection{Rediscretization}
\begin{figure}[ht]
\begin{center}
	\includegraphics[width=0.8\textwidth]{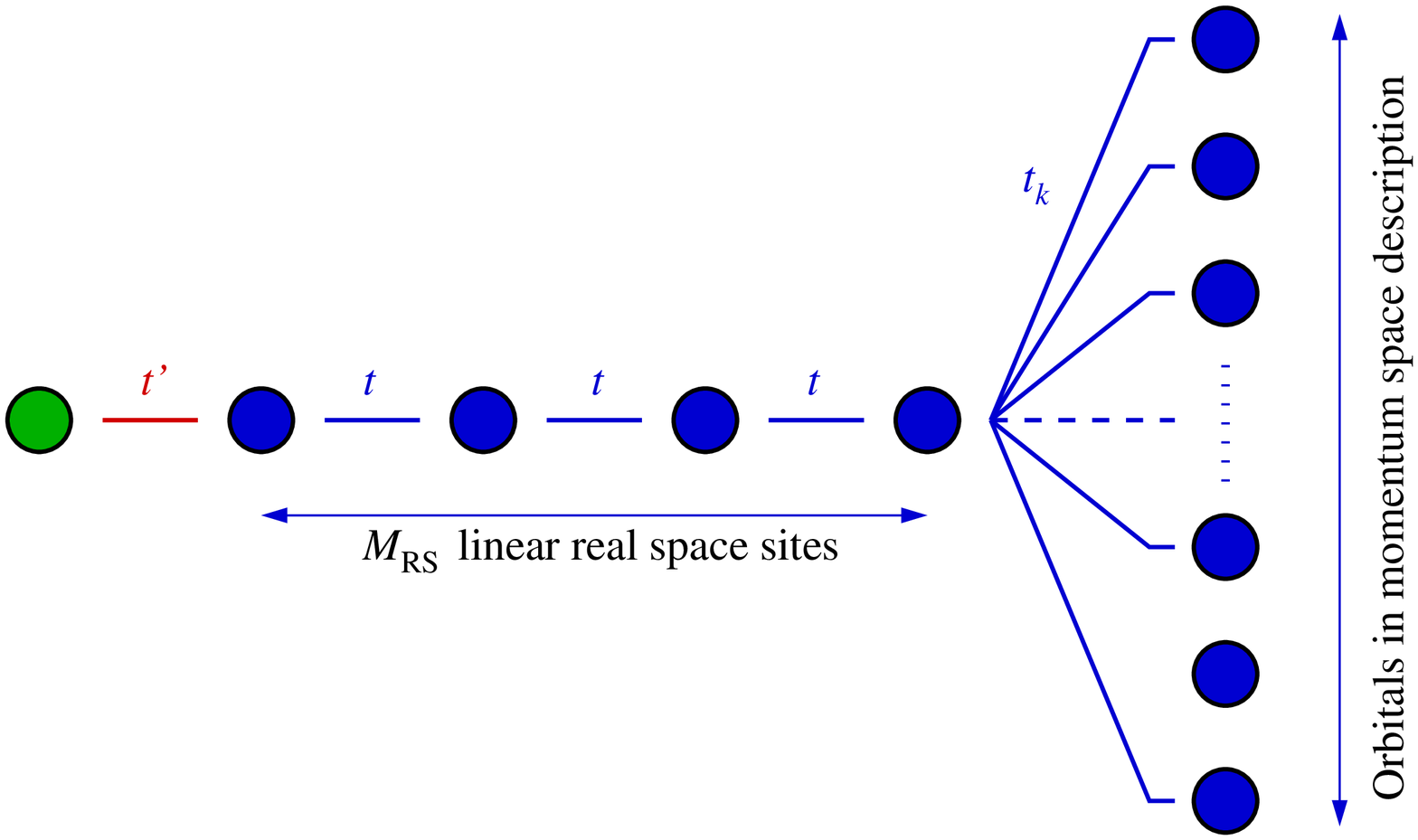}
\end{center}
\Caption{Sketch of a single impurity coupled to $\MRS$ real space sites which are then coupled to a lead in momentum space representation.} \label{fig:CF:NEFS:Fig_SF_ML}
\end{figure}

%%%%%%%%%%%%%%%%%%%%%%%%%
%%
In order to use the leads in momentum space for our numerics we have to rediscretize the leads,
\begin{align}	 
	\HH_{\mathrm{Lead}} &= \sum_{\ell=1}^{\ML} \epsilon^{}_{k_\ell}\, \check{c}^{+}_{\ell} \check{c}^{}_{\ell} \,,
\end{align}
where $k_\ell$ are the \ML\ discretization points between \DL\ and \DU with $k_{\ell-1} < k_\ell$,
which could be used in a scheme as displayed in \Figref{fig:CF:NEFS:Fig_SF_ML}, and $\check{c}^{}_{\ell}$ are the fermionic
anihilation operators in the new discretization scheme.
For convenience we define the interval edges
\begin{align}
	d_\ell &=\; \left\{ \begin{array}{ll} \DL & \ell=0 \\  ( k_\ell + k_{\ell+1}) /2 & 1\le \ell < M\\ \DU & \ell=M \end{array}\right.\,,
\end{align}
the level spacing
\begin{align}
	\Delta_\ell &= d_\ell - d_{\ell-1} \,,
\end{align} 
and density of state weights 
\begin{align}
	\NDOS_\ell &= \int_{ d_{\ell-1}}^{d_\ell} \dd k \,\NDOS(k) \,.
\end{align} 
%%
%%
%%--------------------------------------------------------------------
%%--------------------------------------------------------------------
%%
%%
In order to preserve the density of states
and to ensure canonical commutation relations we have to use the following rule:
\begin{align}
	\frac{1}{\sqrt{2\pi}} \int_{ d_{\ell-1}}^{d_\ell} \dd k \,\sqrt{\NDOS(k)}\,\tilde{c}(k) &\rightarrow\; \sqrt{ \frac{\NDOS_\ell}{2\pi} } \,\check{c}_\ell \,,
\end{align}
which leads to
\begin{align}
	\left[ \check{c}_\ell^{+}, \check{c}^{}_{\ell'} \right]_+ &= \frac{{2\pi}}{\sqrt{\NDOS_\ell \sqrt{\NDOS_{\ell'}}}} \int_{ d_{\ell-1}}^{d_\ell} \frac{\dd k}{\sqrt{2\pi}} \sqrt{\NDOS(k)} \int_{ d_{\ell'-1}}^{d_\ell'}  \frac{\dd q}{\sqrt{2\pi}} q \sqrt{\NDOS(q)}
		\left[ \hat{c}_k^{+}, \hat{c}^{}_{q} \right]_+ \\
   	&= \delta_{\ell,\ell'} \NDOS^{-1}_\ell \int_{ d_{\ell-1}}^{d_\ell} \dd k \, \sqrt{\NDOS(k)} \int_{ d_{\ell-1}}^{d_\ell} \dd q \, \sqrt{\NDOS(q)} \delta_{k,q}\\
   	&= \delta_{\ell,\ell'} \NDOS^{-1}_\ell \int_{ d_{\ell-1}}^{d_\ell} \dd k  \,\NDOS(k)\\
	&= \delta_{\ell,\ell'} \,.
\end{align}
Note that in the discretization of a single onedimensional lead we have $\NDOS(k)=1$ and therefore  $\NDOS_\ell = \Delta_\ell$.
%%
%%
%%%%%%%%%%%%%%%%%%%%%%%%%
\subsubsection{Level Distribution Function}
%%%%%%%%%%%%%%%%%%%%%%%%%
%%
In order to obtain a distribution similar to the logarithmic distribution in section \ref{sec:CF:NEFS:LD} we use
an integrated distribution function for the levels and discretize the interval $[ \DL, \DU]$ in equal area sections.
% % % Here we use the sum of a constant integrated distribution
% % % \begin{align}
% % % 	P_{\mathrm{const}}( p, \DL, \DU ) &=&  \frac{ \Theta(p -\DL) \Theta(\DU - p ) }{ \DU - \DL}
% % % \end{align}
Here we use  a regularized $1/x$ function
\begin{align}
	\tilde{P}_{\mathrm{log}}( p, w_1, w_2) ) &= \left\{ 
		\begin{array}{ll}  
			\frac{1}{\sqrt{ (p-w_2)^2 + w_1^2}} &: p<-w_2\\
			\frac{1}{\sqrt{ w_2^2 + w_1^2} }    &: -w_2 \le p \le w_2\\
			\frac{1}{\sqrt{ (p+w_2)^2 + w_1^2}} &: p>w_2
		\end{array} \right. \\
	 P_{\mathrm{log}}( p, w_1, w_2, \DL, \DU ) ) &= \frac{ \tilde{P}_{\mathrm{log}} ( p, w_1, w_2) ) }{ \int_{\DL}^{\DU} \dd k \,  \tilde{P}_{\mathrm{log}} ( k, w_1, w_2) ) } \,.
\end{align}
to generate the levels.
For $w_2=0$ one obtains a logarithmic distribution for the levels $\epsilon_p$ from $P_{\mathrm{log}}( p, w_1, 0, \DL, \DU ) )$ which
is slightly smoothed on a scale $w_1$. The inset of the constant part in the centre enables a linear spacing in the high resolution
region. This is  important for obtaining accurate results in the calculation of linear transport from the Kubo formula.\cite{BohrSchmitteckert:PRB2007}
%%
%%
% % % %%%%%%%%%%%%%%%%%%%%%%%%%
% % % \subsubsection{Lanczos tridiagonlization}
% % % %%%%%%%%%%%%%%%%%%%%%%%%%
% % % %%
% % % %%
% % % At this point we can make a connection to the form of logarithmic discretization used in NRG, compare section \ref{sec:CF:NEFS:LD}.
% % % Starting from any discretization scheme we can obtain a usual tight binding chain via Lanczos tridiagonalization.
% % % Assuming that our leads are only coupled to a single site of a nano structure we can diagonalize the system under
% % % the restriction that the impurity site remains unchanged, which can for example be achieved by an Arnoldi approach
% % % with modfied Gram-Schmidt orthogonalization.
%%
%%
%%%%%%%%%%%%%%%%%%%%%%%%%
\subsubsection{RLM Spectral Function}
%%%%%%%%%%%%%%%%%%%%%%%%%
%%
For a symmetric dispersion $\epsilon(p) = \epsilon(-p)$ we can now employ another transformation
by taking a symmetric level distribution for  left ($p<0$) and right ($p>0$) movers and combining them to
\begin{align}
	\check{c}_{\pm,p} &= \left( \check{c}_p \pm   \check{c}_{-p} \right) / \sqrt{2}  \qquad p > 0 
\end{align}
leading to the lead Hamiltonians
\begin{align}
	\HH_\pm &=  \sum_{p>0} \epsilon(p) \, \check{c}^+_{\pm, p} \check{c}^{}_{\pm, p} \,,
\end{align}
where only $\HH_+$ couples to the impurity. Therefore we can ignore the $\HH_-$ part of the Hamiltonian.
If we denote the last real space site with $n$, then the coupling to the momentum leads is given by
\begin{align}
	-t \,\hat{c}^+_n \hat{c}^+_{n+1} &\rightarrow -\sqrt{2} t \,\hat{c}^+_n \sum_\ell \sqrt{\NDOS_\ell}\, \check{c}_{+,\ell} \,.
\end{align}
\par
%%----------------------------------------------
\begin{figure}[ht]
\begin{center}
\includegraphics[width=0.9\textwidth]{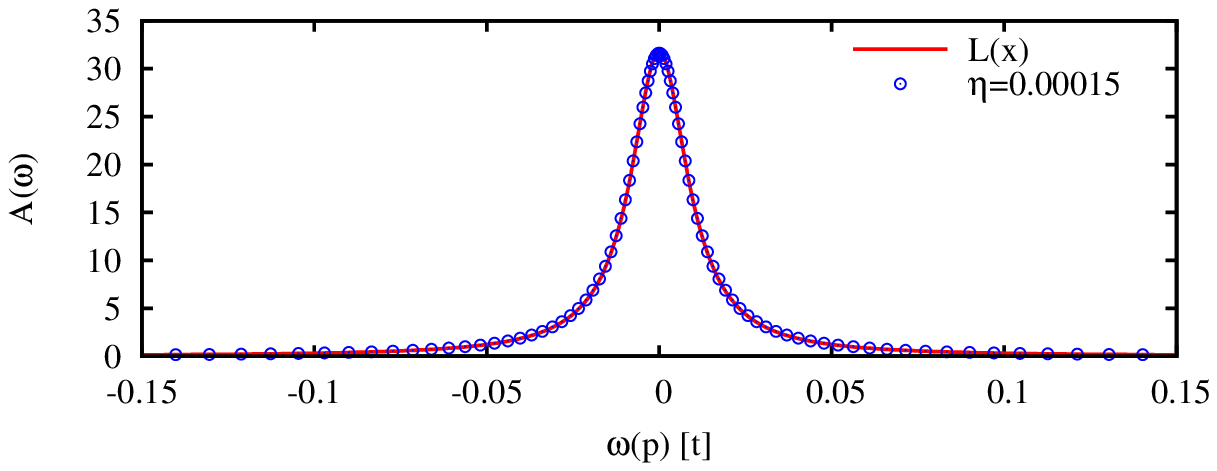}
\end{center} %0.00996
\Caption{Spectral function of a resonant level using momentum leads and frequency adapted grids.
In the numerics we employed a fine grained resolution at $\omega$
using $P_{\mathrm{log}}( p=\omega/2, 0.0001, 0.0002, -1, 1 )$, a linearized dispersion $\varepsilon_k = 2 k t$,
and a constant broadening $\eta=0.00015t$.
The momentum leads consist of 50 sites, and the impurity is first coupled to 3 real space sites.
The full red line shows the Lorentzian  of half halfwidth $w=0.01$ corresponding to the spectral function
in the wide band limit.} \label{fig:CF:NEFS:SF_ML_FAG}
\end{figure}
%%--------------------------------------------------------
The main advantage of momentum leads is that one can adapt the discretization to the frequency $\omega$
that should be resolved. We test this idea by generating a small level distance at frequency 
$\omega=-2t\cos(p)$ by slicing $P_{\mathrm{log}}( p, w_1, 2 w_1, \DL, \DU)$ and plot the result in \Figref{fig:CF:NEFS:SF_ML_FAG}.
Clearly, this approach gives an excellent result which reproduces the central peak and
the tails. In addition,
we can even go back to use a fixed $\eta$, as the level spacing at $\omega$ is now always of the same order of magnitude.
This allows us to fit \Eqref{eq:CF:ConvolutedLorentzian} leading to a bare width of $w_{\mathrm{ML}}=0.00991$.
In addition, the momentum lead approach allows us to restrict the bandwidth to the relevant region.
Keeping the parameter of \Figref{fig:CF:NEFS:SF_ML_FAG} and only changing the momentum cutoff to $\pm 0.1$
we obtain a bare width of $w_{\mathrm{ML}}=0.00997$.
\par
\begin{figure}[ht]
\begin{center}
\includegraphics[width=0.9\textwidth]{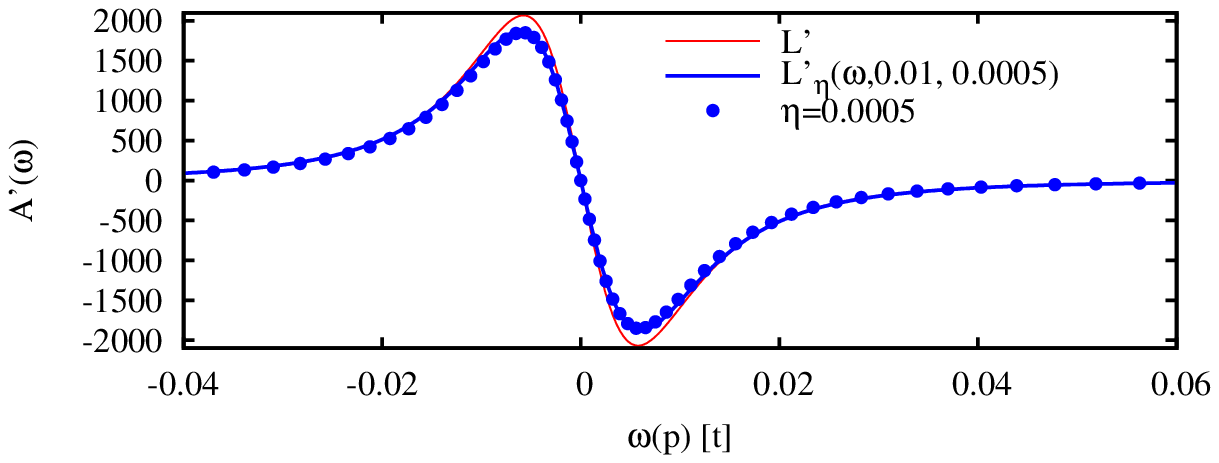}
\end{center}
\Caption{Derivative of the spectral function of a resonant level using momentum leads.
The red line is the analytical result in the wide band limit, the blue line is
the analytical result convoluted with a Lorentzian of half halfwidth $\eta=0.0005$.
The circles are obtained from momentum leads using a distribution centred around $\pm \omega$ with
$P_{\mathrm{log}}( p, w_1, 2 w_2)$, $w_1=0.0005$, $w_2=0.0001$,
a linear lead $\omega(k) = 2 k t$  with cutoffs $D_{\mathrm{l,u}}=\pm 0.1t$, and 3 real space sites
sandwiched between the impurity and the momentum leads.} \label{fig:CF:NEFS:SF_ML_Diff}
\end{figure}

Finally we demonstrate in \Figref{fig:CF:NEFS:SF_ML_Diff} that within this scheme one can even
obtain accurate derivatives of Green functions with respect to the frequency.
The derivative of $\GG^\pm$  is given by
\begin{align}
	\frac{ \dd }{ \dd \omega} \GG^+_{\hat{A},\hat{B}}(\omega)&=
	  \bra{\Psi_0}  \hat{A} \,\frac{-1}{ \left( E_0 - \HH + \omega + \im\eta\right)^2 }  \, \hat{B} \ket{\Psi_0}  \label{eq:CF:Gp_Diff}\\
	\frac{ \dd }{ \dd \omega} \GG^-_{\hat{A},\hat{B}}(\omega)&=
\bra{\Psi_0}  \hat{A} \,\frac{1}{ \left(E_0 - \HH - \omega - \im\eta \right)^2}  \, \hat{B} \ket{\Psi_0} \,.\label{eq:CF:Gm_Diff}
\end{align}
It turns out that the numerical evaluation of \Eqsref{eq:CF:Gp_Diff}{eq:CF:Gm_Diff} is more sensitive to
the discretization used. We switch to a linear band, $\omega=2kt$, with cutoffs of $D_{\mathrm{l,u}}=\pm 0.1$,
100 momentum lead sites, the distribution function $P_{\mathrm{log}}( p, w_1, 2 w_1)$, $w_1=0.0005$, 
and $\eta=0.0005$.
As a comparison we plot the exact result and the exact result convoluted with a Lorentzian of half halfwidth $\eta=0.0005$,
\begin{equation}
	 L'_\eta(\omega,w) = \frac{ \dd }{ \dd \omega} L_\eta(\omega,w) = 
	 \frac{-2\omega}{\pi} \frac{ w + \eta }{ \left( \omega^2  + ( w + \eta)^2 \right)^2 } \,.\label{eq:CF:ConvolutedLorentzian_Diff}
\end{equation}
By fitting $L'_{0.0005}(\omega,w)$ of \Eqref{eq:CF:ConvolutedLorentzian} to the numerical result we obtain $w=0.01004$.
For the same discretization a fit \Eqref{eq:CF:ConvolutedLorentzian} of the spectral function gives $w=0.0099996$.
%%
%%----------------------------------------------
%%--------------------------------------------------------
%%
%%
%%%%%%%%%%%%%%%%%%%%%%%%%
\section{Bulk Green Functions}
\label{ssec:CF:NEFS:BGF}
%%%%%%%%%%%%%%%%%%%%%%%%%
%%
In the previous sections we discussed the spectral properties of an impurity, namely a  single resonant level,
coupled to a non-interacting lead. In this case it was natural to look at the local spectral functions of the impurity.
We now turn to the momentum resolved spectral function of a bulk system. As an example we look at the 
retarded Green function of a tight binding chain
\begin{align*}
  \HH &= - t \sum_x \hat{c}^+_{x} \hat{c}_{x-1} \,+\, \hat{c}^+_{x-1} \hat{c}_{x} = -2 t \int_{-\pi}^{\pi} \dd k \, \cos(k) \, \hat{f}^+_{k} \hat{f}_{k} \\
    \hat{f}_{k} &=  \frac{1}{\sqrt{2\pi}} \sum_x \e^{ \im k x } \, \hat{c}_{x} 
\end{align*}
At $k=\pi/2$ the spectral function is simply given by  $A(\omega) = \delta(\omega)$.
\subsection{Choosing Correct Single Particle States}
For simplicity let us start with a sine function ansatz for the single particle states on $M$ sites
\begin{align}
    \hat{f}_{n} &=  \frac{1}{\sqrt{M}} \sum_{x=1}^{M} \sin( k_n x ) \, \hat{c}_{x}  \label{eq:SineAnsatz1}\\
    k_n &= \frac{ 2 \pi n }{ M }  \label{eq:SineAnsatz1_k}
\end{align}

\begin{figure}[ht]
\begin{center}
\includegraphics[width=.9\textwidth]{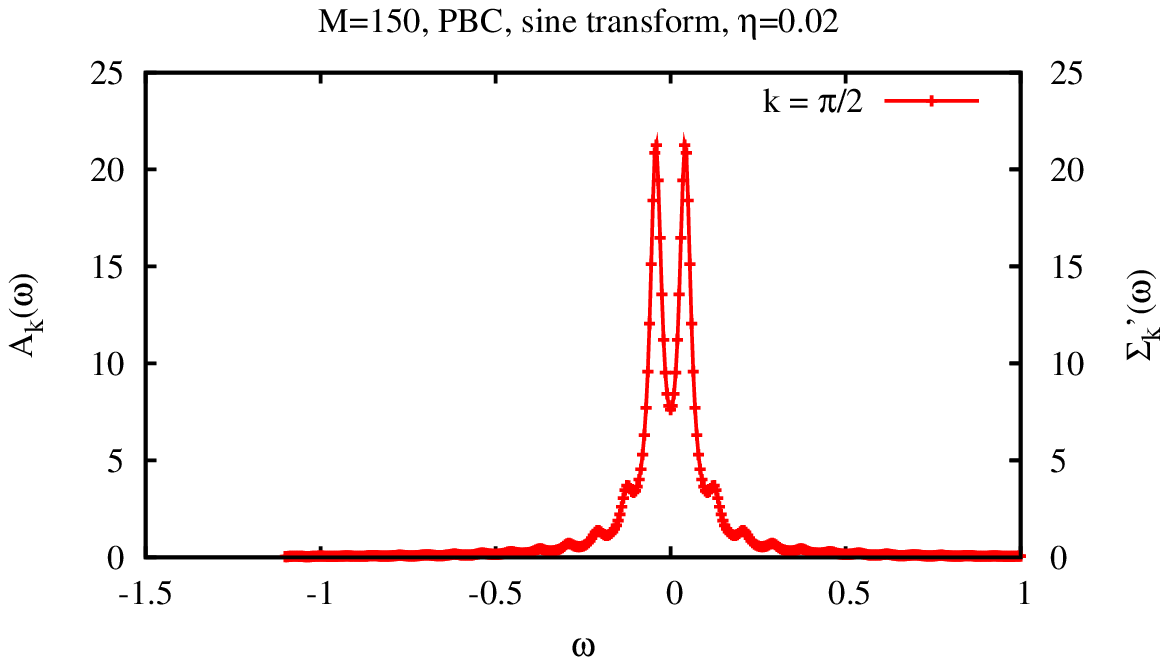}
\end{center}
\Caption{Spectral function of the tight binding chain at  $k=\pi/2$ using the single particle states of \Eqsref{eq:SineAnsatz1}{eq:SineAnsatz1_k}
and a tight binding chain with $M=150$ lattice sites and periodic boundary conditions (PBC)  and sine function as the basis set.} \label{fig:Bulk_PBC}
\end{figure}

As can be seen in \Figref{fig:Bulk_PBC} this ansatz leads to a fictitious double peak structure and an oscillatory part which
does not resemble the $\delta$-peak of the system in the thermodynamic limit. The reason for this is that by using the single
particle states of \Eqsref{eq:SineAnsatz1}{eq:SineAnsatz1_k} we used single particle states which are not eigenstates of the system
with periodic boundary conditions (PBC).
\par
\begin{figure}[ht]
\begin{center}
\includegraphics[width=.9\textwidth]{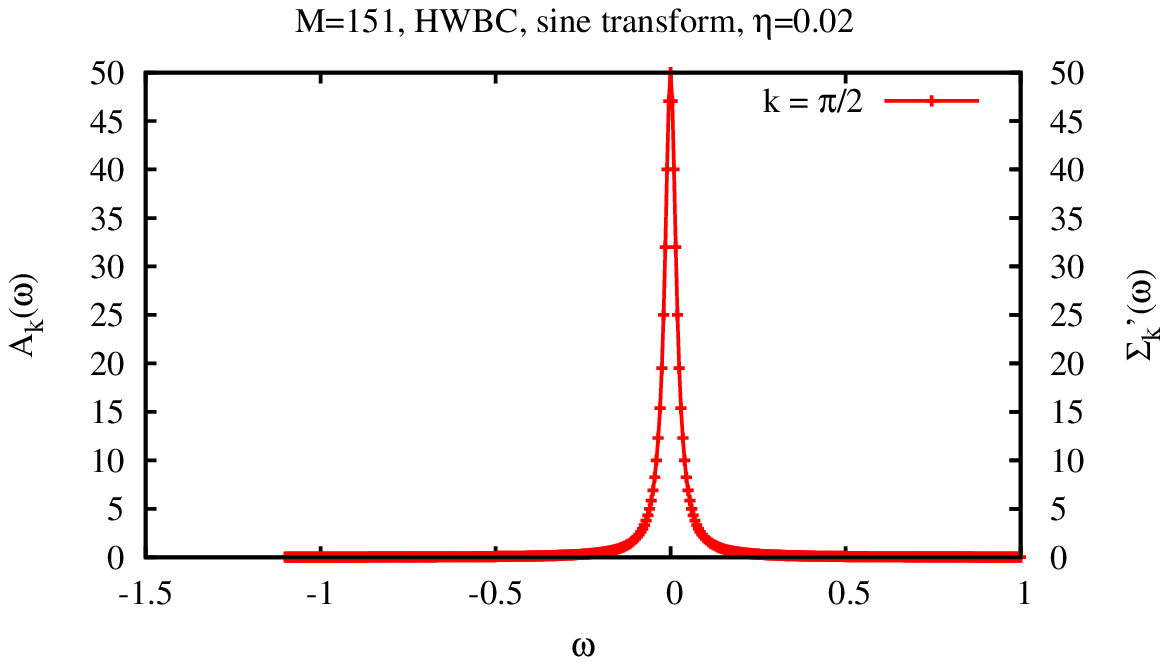}
\end{center}
\Caption{Spectral function of the tight binding chain at  $k=\pi/2$ using the single particle states of \Eqref{eq:SineAnsatz}
and a tight binding chain with $M=150$ lattice sites and periodic boundary conditions (PBC) and sine function as the basis set.} \label{fig:Bulk_HWBC-2}
\end{figure}

Therefore we repeat this calculation for a tight binding chain employing hard wall boundary conditions (HWBC).
\begin{figure}[ht]
\begin{center}
\includegraphics[width=.9\textwidth]{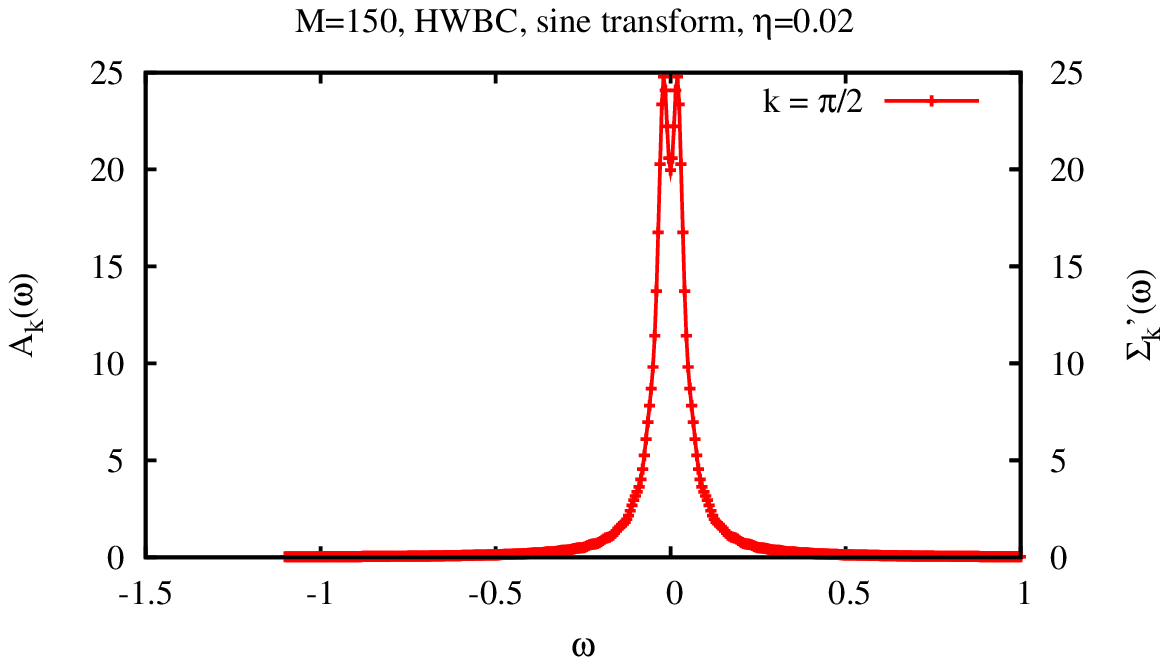}
\end{center}
\Caption{Spectral function of the tight binding chain at  $k=\pi/2$ using the single particle states of \Eqsref{eq:SineAnsatz1}{eq:SineAnsatz1_k}
and a tight binding chain with $M=150$ lattice sites and periodic boundary conditions (PBC) and sine function as the basis set.} \label{fig:Bulk_HWBC}
\end{figure}
The graph in \Figref{fig:Bulk_HWBC} shows that the result looks much better now, but we still obtain the fictitious double peak structure.
Again, the reason for this lies in a wrong single particle basis. However, this time the reason for its failure it is more subtle.
The problem is that \Eqref{eq:SineAnsatz1_k} gives the wrong eigenstates for HWBC and the correct single particle basis is given by
\begin{equation}
    \hat{f}_{n} =  \sqrt{\frac{2}{{M+1}}} \sum_{x=1}^{M} \sin( k_n x ) \, \hat{c}_{x}  \qquad k_n = \frac{ n \pi  }{ M+1 } \quad  n = 1, 2, \cdots, M \label{eq:SineAnsatz}
\end{equation}
leading to the result of \Figref{fig:Bulk_HWBC-2} which finally resembles the $\delta$-peak broadened by $\eta$.
In summary we would like to point out that one can obtain nice spectral functions for the bulk system from finite system, however
care has to be taken to choose the correct representation. Otherwise spurious structures may appear.
The need for using the sine solution for HWBC has been pointed out by Benthien, Gebhard, and Jeckelmann\cite{Benthien_Gebhard_Jeckelmann:PRL2004}.
It was shown by Ulbricht and Schmitteckert that for interacting particle in a harmonic trap one can obtain spectral functions
from finite system by resorting to hermite polynomials.\cite{Ulbricht_Schmitteckert:EPL2010}
\section{Poor Man's Deconvolution: Self Energy Sharpening}
In this section we discuss a simple strategy to ``sharpen'' Green functions obtained in the previous sections.
It was first applied to improve the spectral function of a spin polarized ondedimensional Hubbard model.\cite{Ulbricht_Schmitteckert:EPL2010}
As an example we improve the result for the spectral function of the tight binding chain of the previous section.
A straight forward strategy to improve the result consists in a deconvolution to remove the broadening introduced by the
finite $\eta$ in the denominator of the resolvents, i.e.\ perform the inverse operation of \Eqref{eq:CF:ConvolutedLorentzian}.
While this procedure is mathematically well defined, it is highly unstable numerically and very sensitive to numerical noise.\cite{Raas_Uhrig:EPJB2005}
Here we want to introduce a simple method to improve the results, which has the advantage of being stable
and allows to preserve analytical properties of spectral functions, e.g. $ \A (\omega) \le 0$.

\begin{figure}[ht]
\begin{center}
\includegraphics[width=.8\textwidth]{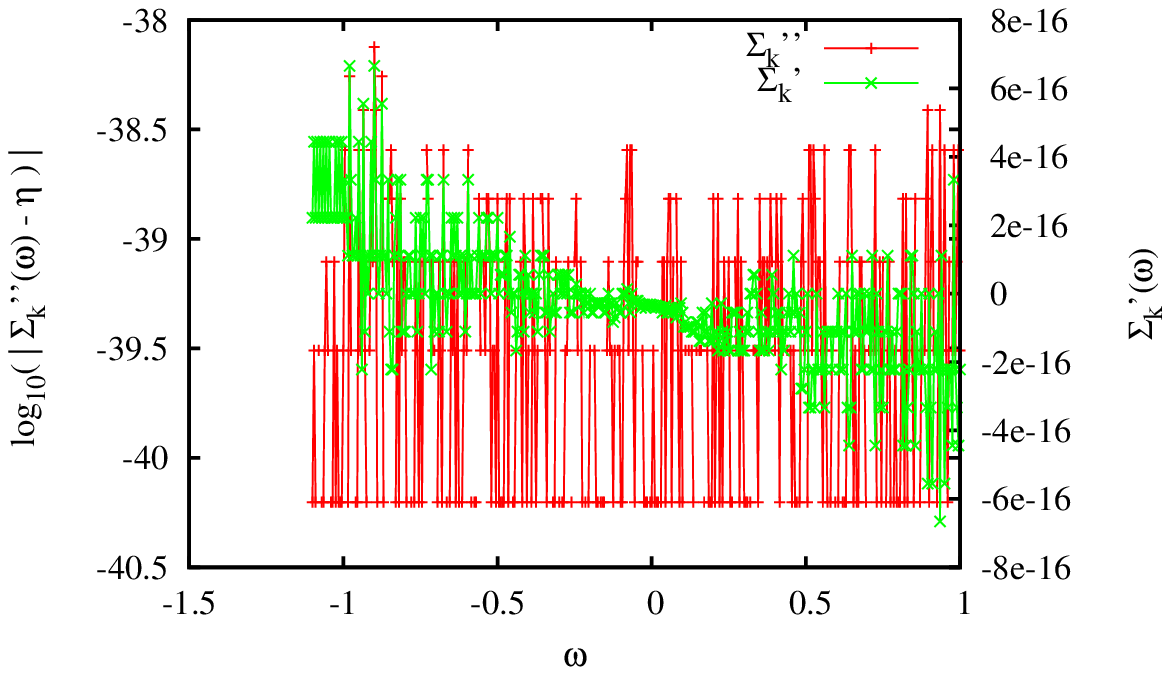}
\end{center}
\Caption{Imaginary ($\Sigma''$) and real ($\Sigma'$) part of the self energy of tight binding chain obtained by sharpening the results \Figref{fig:Bulk_HWBC-2}
for a 151 site tight binding chain at  $k=\pi/2$ using the single particle states of \Eqref{eq:SineAnsatz},
hard wall boundary conditions (HWBC) and sine function as the basis set.} \label{fig:Bulk_DS}
\end{figure}

We start by defining the self energy $\Sigma(\omega)$ for the infinite system
\begin{equation}
	\GG^r(\omega) = \frac{ 1}{\omega - \Sigma(\omega) + \im0^+} 
\end{equation}
and the self energy $\Sigma_\eta(\omega)$ for the finite system which includes the finite broadening $\eta$
\begin{equation}
	\GG^r_\eta(\omega) = \frac{ 1}{\omega - \Sigma_\eta(\omega) + \im0^+} \,.
\end{equation}
When switching from the infinite system to the finite system we should trace out the discarded part which 
would lead to a contribution to the self energy of the reduced system. However, we do not know this part.
Therefore we make the ansatz that the discarded part can be modeled by the $\im\eta$ term we already have in the 
resolvents to enable the mixing of excited states. In this way we obtain our
``poor man's deconvolution'' ansatz $\Sigma_\eta = \Sigma + \im\eta $:
\begin{align}
	 \Sigma_\eta(\omega) &= \omega -  \frac{ 1}{\GG^r_\eta(\omega) } + \im0^+ \\
	 \Sigma(\omega)      &= \omega -  \frac{ 1}{\GG^r_\eta(\omega) } - \im\eta + \im0^+
\end{align}
where $\GG^r_\eta(\omega) $ is the Green function we obtain from our numerics on a finite lattice
and $\eta$ is the broadening explicitly used in the numerics.
The result of this sharpening applied to the data of \Figref{fig:Bulk_HWBC-2} is shown in  \Figref{fig:Bulk_DS}. Since
real and imaginary part are vanishing up to numerical precision we have rediscovered the $\delta$-peak structure of the infinite system
from the numerics on a finite lattice.

%%%%%%%%%%%%%%%%%%%%%%%%%%%%%%%%%%%%%%%%%%%
\section{The NRG Tsunami}
%%%%%%%%%%%%%%%%%%%%%%%%%%%%%%%%%%%%%%%%%%%

\begin{figure}[ht]
\begin{center}
\includegraphics[width=.7\textwidth]{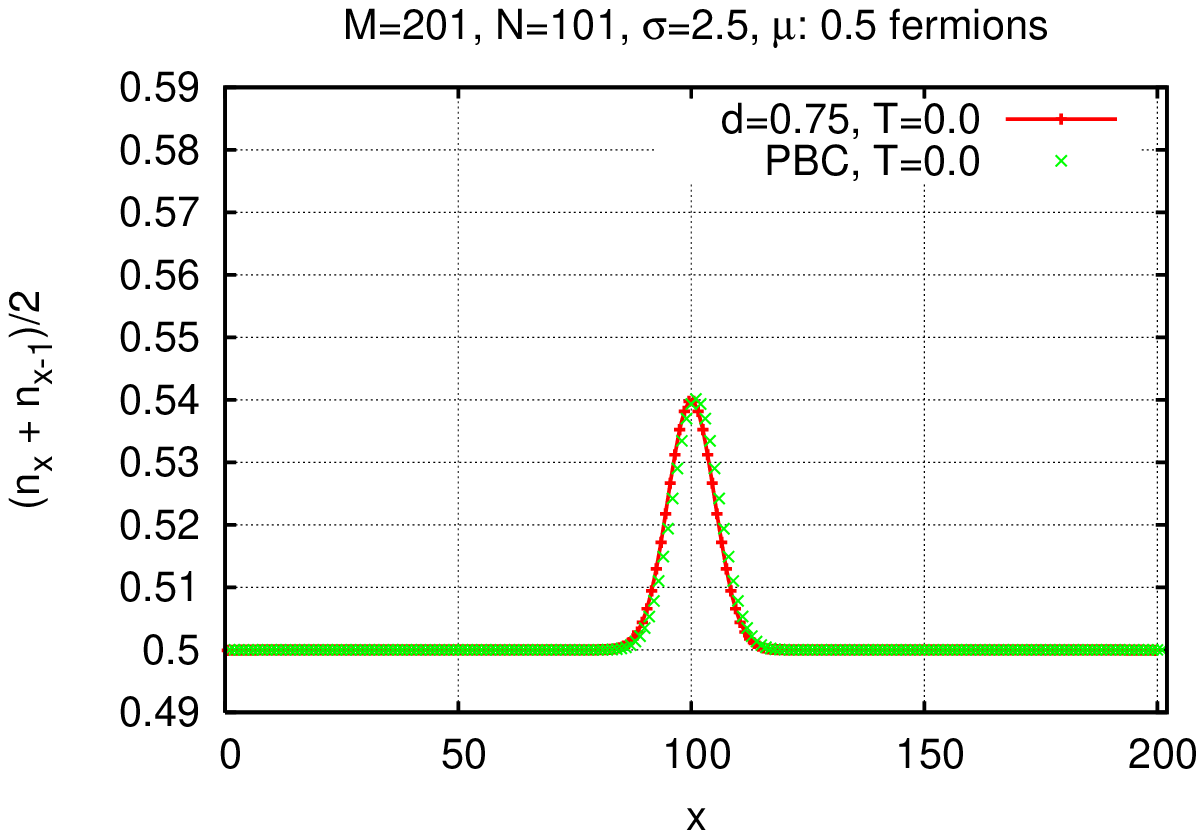}
\end{center}
\Caption{Density of a $M=201$ site tight binding chain with $N=101$ fermions, where a Gaussian potential of width $\sigma=2.5$ was applied to trap an additional
half fermion in the centre region in addition to an average density of $\rho=0.5$. In the results for periodic boundary conditions (PBC) are given by the crosses.
The result for Damped Boundary conditions as displayed in \Figref{fig:CF:NEFS:Fig_SF_DBC} applied to the left and right end of the chain is displayed by the
line with plusses.} \label{fig:Tsunami-0}
\end{figure}

Finally we would like to discuss the effect of damped boundary conditions on the dynamics of wave packets to show
that boundary conditions are not  only important in frequency space, but that they can also change the results in time domain dramatically.
In this section we follow the route of \cite{Schmitteckert:PRB2004} adapted for free fermions using exact diagnalization
of the quadratic form.
In \Figref{fig:Tsunami-0} we show the initial states of a 201 site tight binding chain with periodic boundary conditions at half filling, where
half a fermion was trapped in the middle of the system by applying a Gaussian potential with a width of $\sigma=2.5$.
In addition we show the response to the same perturbation for a system where the PBC are replaced by damped boundary conditions,
compare \Figref{fig:CF:NEFS:Fig_SF_DBC}.

\begin{figure}[ht]
\begin{center}
\includegraphics[width=.7\textwidth]{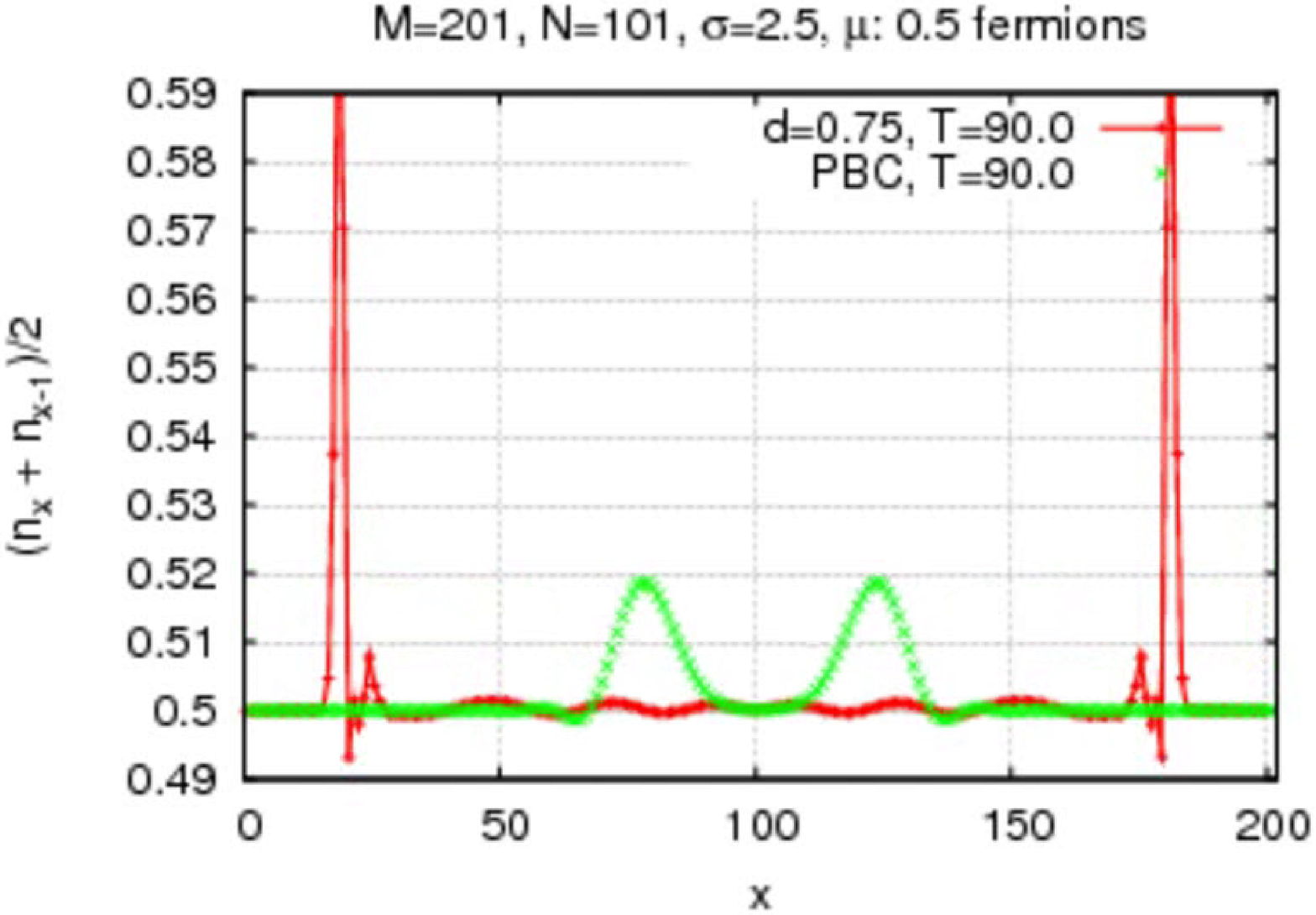}
\end{center}
\Caption{Density of a $M=201$ site tight binding chain with $N=101$ fermions, where a Gaussian potential of width $\sigma=2.5$ as in \Figref{fig:Tsunami-0}
after performing a time evolution of time $T=90$ with the Hamiltonian without the perturbation. The results for periodic boundary conditions (PBC) are given by the crosses.
The result for damped boundary conditions as displayed in \Figref{fig:CF:NEFS:Fig_SF_DBC} applied to the left and right end of the chain is shown by the
line with plusses.} \label{fig:Tsunami90}
\end{figure}

In \Figref{fig:Tsunami90} we show the system after performing a time evolution up to time $T=90$, in which the homogeneous system moves a distance
of $T * \vF=180$ sites. Due to the periodic boundary conditions the wave packets appear now at position $x \approx 100 \pm 20$.
Note that by applying a low energy perturbation we create excitations around $\pm \kF$ and our initial wave
packet splits into a right and left moving wave packet travelling at $\pm \vF$, where $\kF$ is the Fermi momentum and $\vF$ is the
excitation velocity. However, once the wave packet hits the region of damped boundary conditions (DBC) it sees an exponentially decreasing
hopping element which results in an exponentially decreasing excitation velocity. In the similar way as long water waves hitting the shore,
where the excitation velocity is decreased, our wave packet has to pile up like a tsunami, since the front is moving much slower than the back.
In addition each changed hopping element creates a small back reflection leading to an additional wiggling.

%%%%%%%%%%%%%%%%%%%%%%%%%%%%%%%%%%%%%%%%%%%
\section{Summary}
%%%%%%%%%%%%%%%%%%%%%%%%%%%%%%%%%%%%%%%%%%%
In summary we have shown that one can extract sound results for the thermodynamic limit in the context of calculating Green functions
from a finite lattice. However, care has to be taken to choose the correct representation of states.
In order to calculate spectral functions at finite frequencies for impurity problems one should apply
frequency adapted grids to achieve high resolution. A sharpening procedure based on the self energy can be used as a replacement for
deconvoluting the broadening introduced by a finite $\eta$.

%%%%%%%%%%%%%%%%%%%%%%%%%%%%%%%%%%%%%%%%%%%
%{acknowledgement}
%%%%%%%%%%%%%%%%%%%%%%%%%%%%%%%%%%%%%%%%%%%
\acknowledgments
%\begin{acknowledgements}
%\section*{Acknowledgments}
The author wishes to acknowledge assistance and clarifying discussions
with Alexander Branschädel, Dan Bohr, Stefan Kremer, Christina Stawiarsky, and Tobias Ulbricht.
%\end{acknowledgements}

%%%%%%%%%%%%%%%%%%%%%%%%%%%%%%%%%%%%%%%%%%%
%\section{References}
%%%%%%%%%%%%%%%%%%%%%%%%%%%%%%%%%%%%%%%%%%%
%%\bibliographystyle{unsrt}
%%\bibliography{../IOP}

\end{document}